# Historical overview on Vacuum suitable Welding and fatigue resistance in Research Devices

Martin Wolf  E-mail:emartin.wolf@gmail.com

## Summary

New inventions change the approach of vacuum suitable welding for research purpose. With orbital welding, laser welding and robot welding the possibilities increase to fabricate larger vessels more accurately. Despite this development there is still no perfect understanding on how to avoid virtual leaks and how to make such joints suitable for dynamic stress. By recalling its historical development, it is apparent how welding mistakes began occurring systematically and how to avoid them. With ASDEX-Upgrade as an example, it is shown how the attempt to conduct vacuum suitable welding has decreased the fatigue strength. ITER could repeat the mistakes of ASDEX-Upgrade even for unwanted welding (accidental fusing of joints).

## Virtual Leak

When vacuum chambers started to be larger and larger, there was the need to change from glass chambers to welding fabricated stainless steel vessels. Therefore welding had to meet vacuum requirements. With methods like helium leak detection (HLD) available it took some effort to check the welded chambers and to get welding mistakes repaired. It should be mentioned that HLD is a very accurate method, capable of finding mistakes that X-ray, ultrasonic and dye penetrant inspection cannot find.

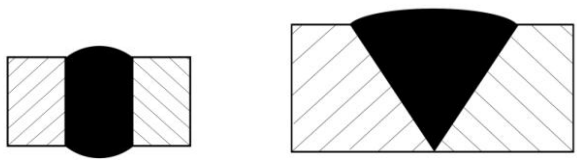

Figure 1: I-Seam and V-Seam both one Layer

Vessels were fabricated with I-seams or V-seams applied from outside (Fig.1). The weld was a single pass. The vacuum performance was high-quality with little impact from virtual leaks of the welding. With the demand for even larger chambers, X-seams consisting of two welded layers were applied to assemble sheets of greater thickness.

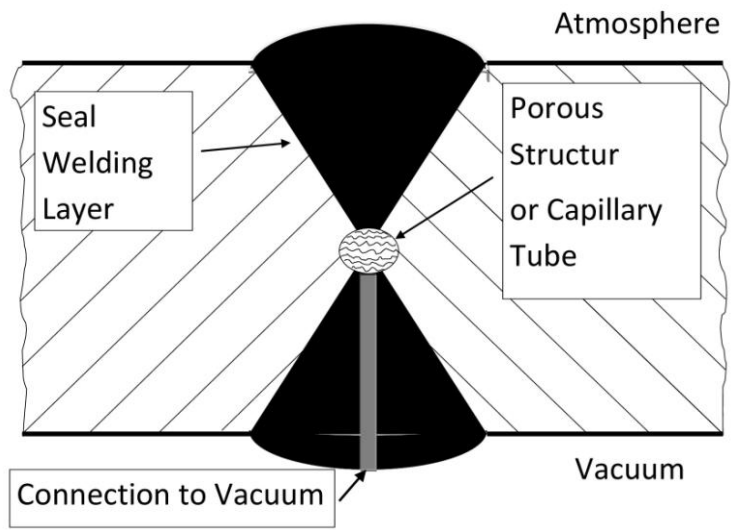

**Figure 2: X-Seam with virtual Leak (schematically).**

The X-seam welding is susceptible to the creation of trapped volumes, if the weld penetration is not good enough a capillary tube is created at the X-point region. In addition, at the X-point region, the welding conditions are the worst, where the content of base metal in the welding seam is at its maximum and the cooling velocity is fastest, therefore the conditions for creating a porous structure are optimal. Thus, even if a capillary tube is avoided by good penetration, there can be still a trapped porous volume with a connection to the vacuum chamber (Fig.2). The vessel will then suffer from virtual leak. This means the pumping down takes more time, leaving less time for experiments and less plasma purity. The measure being taken at DESY was of the order: Welds must be completed on the vacuum side of the vessel only [1]. At ADEX-Upgrade it is the same praxis: welding only from the inside. Figure 3 shows a welded feed through pipe at ADEX-Upgrade, sealed from the inside (schematically). It is clearly visible that the structure with many weld passes offers multiple possibilities for virtual leaks. Similar to the X –seam, there can be lack of side wall fusion. In addition it is possible to have incomplete interpass fusion, unremoved silica lines and tungsten traces create hidden volumes and connect them.

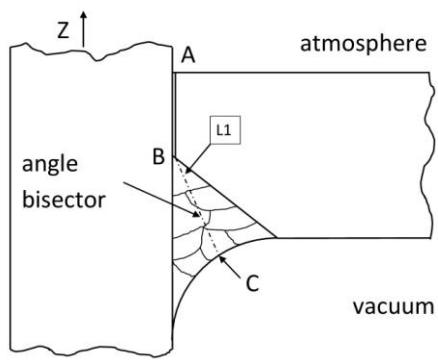

**Figure 3: Welded Pipe Feed Through at ASDEX-Upgrade.**

Even if the initial run (L1) is sealed vacuum tight, the additional layers offer the possibility for all kinds of welding mistakes that could create a labyrinth of connected volumes acting as virtual leak.

# Vacuum suitable welding

To obtain a vacuum sealed joint, only one welding run has to be done on the inside. Structural runs have to be done outside with short skip runs that cover 30-45% of the joint extent so as not to disturb the HLD. Figure 4 shows a vacuum suitable joint for a welded feed through pipe.

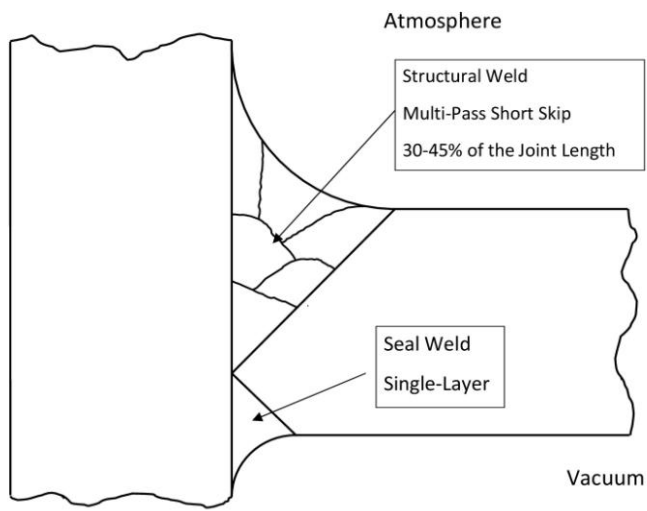

Figure 4: Welded Pipe Feed Through vacuum suitable and low stress concentration.

# Cyclic Loading and Stress

In a state of the art magnetic confinement fusion device, it is common to have recurring Lorentz forces due to sinusoidal, cyclic pulse and shock event electric currents. Due to plasma ignition and vertical plasma movement electric current is induced in passive loops that create vertical forces. Edge-localized modes give electric current pulses from the plasma to the internal assembly. In-vessel coil with sinusoidal electric current creates vertical forces. Plasma disruption gives strong electrical current from the plasma, which when in contact with the internal assembly then create strong forces. In ASDEX-Upgrade the welding joint in Figure 3 is used for the suspension of the vessel and for feeding through and connecting the load coming from the passive stabilizing loops (PSL) to the vessel and suspension rod. This welding joint leaves an un-welded long gap from A to B. The radius at B is infinitesimal. Consequently, it acts as a strong stress concentration in Point B. Already small cyclic Lorentz forces with a z-component create enough stress to make a crack grow from B to C following about the angle bisector. For the operation of ASDEX-Upgrade this means that as soon as the crack passes the sealing part of the welding joint to be tangent to the virtual leak labyrinth a rather small real leak is created. In the long run, the behavior of the machine gradually increases leakage rate. This effect is already noted at ASDEX-Upgrade and JET [2]. As soon as the crack reaches Point C a big leak opens resulting in too high pressure in the machine to continue plasma operation. This kind of leak is not detectible by HLD because the outside part of the joint is not accessible due to the coil system surrounding the vessel. Therefore the operational stop would be permanent. Due to the inaccessibility of the welding joint it is impossible to find out how far the crack has proceeded. After 24 years of operation, about 30 000 experiments and a countless number of stress cycles it cannot be denied that the next plasma ignition could be the fatal one. There are measures to be taken to prevent the fatality on request from the author available.

The welding joint will have reasonable fatigue resistance due to the filling of the intermittent welds on the outside of the structure upon completion of HLD testing and its notable radii. Vacuum performance is improved as a result of the single internal welding run. Both of which are shown in Figure 4.

# ITER

For ITER the vacuum vessel gravity support is placed far outside of the torus center of gravity of the line, at the end of the lowest horizontal port [3] thus there will be a notable side load with a cyclic component on the welding joints connecting it to the vessel. The main load of the vessel is on nine lower supports creating hyper-static system. Due to the welding distortion, the geometry of the lower ports will be slightly dissimilar, also the static load on the ports will vary. Thus, there will be self-equilibrating stress on the lower port joints. This is not considered in the structural analysis. It is assumed to have a cyclic symmetry in the finite element model [4].

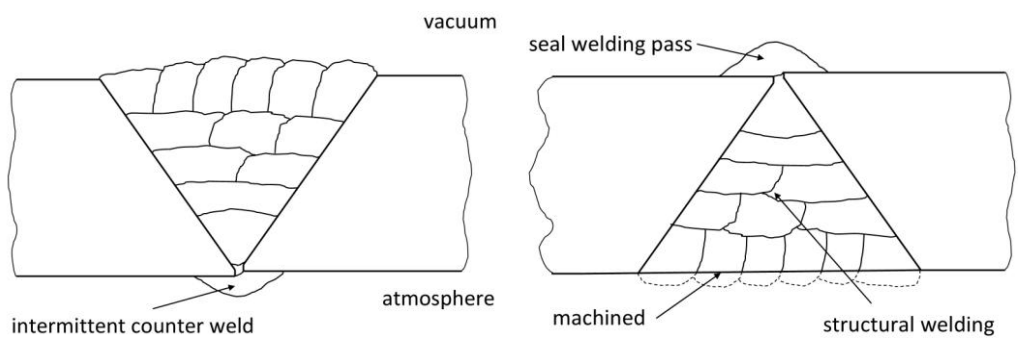

Figure 5:Welding joint atITER with stress concentration (left)and improved (right)

So far there was no stop of operation for a tokamak, due to fatigue strength failure in the main structure. Thus endurance strength is considered irrelevant in ITER structural analysis. Therefore stress concentration in welding joints due to misalignment, root fusion defect and missing or incomplete counter run is ignored in finite element analyses. Figure 5 shows the welding according to the ITER vacuum design handbook [5] with the aforementioned welding mistakes causing stress concentration. Bad vacuum performance comes from multi pass welding inside the vessel (left). A welding joint with a very low stress concentration factor and good vacuum suitability (right) is established by completing the intermittent welding after HLD and machining off the excess weld metal. As mentioned before, the seal welding is single pass for good vacuum performance.

By ignoring self-equilibrating stress and stress concentration in welding joints is to be expected that ITER will face fatigue cracks in cyclic loaded structural welding joints of the lower port.

# Unwanted Welding

For the PSL suspension at ASDEX-Upgrade, electrical isolated hinges are used. Upon inspection after several years of operation it was noticed that the dowel and block welded together at spots causing unwanted plastic deformation when baking. ITER will have hinges without electrical isolation [6] connecting the vessels lower port to the cryostat. This construct is bound to create problems similar or worse to ASDEX-Upgrade when the machine is baked due to blocked thermal expansion.